\documentclass[%
superscriptaddress,
 amsmath,amssymb,
 aps,
twocolumn
]{revtex4-2}

\usepackage{graphicx}
\usepackage{dcolumn}
\usepackage{bm}
\usepackage{xcolor,soul}

\usepackage{hyperref}

\usepackage[normalem]{ulem}
\usepackage{physics}

\DeclareRobustCommand{\vect}[1]{
  \ifcat#1\relax
    \boldsymbol{#1}
  \else
    \mathbf{#1}
  \fi}

  \newcommand{\av}[1]{\left\langle#1\right\rangle}
  \newcommand{\cbr}[1]{\left(#1\right)}

  \newcommand{\new}[1]{#1}

\newcommand{\uv}{\vect{u}}
\newcommand{\fv}{\vect{f}}
\newcommand{\w}{u_1}
\newcommand{\y}{u_2}
\newcommand{\z}{u_3}

\renewcommand{\pv}{\vect{\varphi}}
\newcommand{\pzero}{\varphi_0}

\begin{document}

\title{Optimal control of levitated nanoparticles through finite-stiffness confinement}

\author{Marco Baldovin}
 \email{marco.baldovin@cnr.it}
\affiliation{Institute for Complex Systems, CNR, Universit\`a Sapienza, I-00185, Rome, Italy}
\affiliation{LPTMS, CNRS, Universit\'e Paris-Saclay, F-91405, Orsay, France}
\author{Ines Ben Yedder}
\affiliation{CentraleSup\'elec, LuMIn, Université Paris-Saclay, ENS Paris-Saclay, CNRS, F-91405, Orsay, France}
\author{Carlos A. Plata}
\affiliation{F\'{\i}sica  Te\'orica, Multidisciplinary Unit for Energy Science, Universidad  de  Sevilla, Apartado de Correos 1065, E-41080, Sevilla, Spain}
\author{Damien Raynal}
\affiliation{CentraleSup\'elec, LuMIn, Université Paris-Saclay, ENS Paris-Saclay, CNRS, F-91405, Orsay, France}
\affiliation{Centre of Light for Life (CLL) and Institute for Photonics and Advanced Sensing (IPAS), The University of Adelaide, 5005, Adelaide, SA, Australia}
\author{Lo\"ic Rondin}
\affiliation{CentraleSup\'elec, LuMIn, Université Paris-Saclay, ENS Paris-Saclay, CNRS, F-91405, Orsay, France}
\author{Emmanuel Trizac}
\affiliation{LPTMS, CNRS, Universit\'e Paris-Saclay, F-91405, Orsay, France}
\affiliation{\'Ecole  Normale  Sup\'erieure de Lyon, F-69342, Lyon, France}
\author{Antonio Prados}
\affiliation{F\'{\i}sica  Te\'orica, Multidisciplinary Unit for Energy Science, Universidad  de  Sevilla, Apartado de Correos 1065, E-41080, Sevilla, Spain}

\date{\today}

\begin{abstract}
Optimal control of levitated nanoparticles subjected to thermal fluctuations is a challenging problem, both theoretically and experimentally. In this Letter, we compute the time-dependent harmonic confining potential that steers, in a prescribed time and with the minimum energetic cost, a Brownian particle between two assigned equilibrium states. We take full account of inertial effects, thus addressing the general underdamped dynamics, and, to address actual experimental conditions, the stiffness of the confining potential is required to be bounded. We carry out an experiment realizing the described protocol for an optically confined nanoparticle, which is shown to reach the target state within accuracy---while spending less energy than other protocols with the same duration, significantly shorter than the characteristic relaxation time. The results presented here are expected to have relevant applications in the design of optimal devices, such as engines at the nanoscale. 
\end{abstract}

\maketitle


The field of optimal control aims at devising time-dependent protocols for the parameters controlling the dynamical evolution of a system, while minimizing a certain cost functional of its trajectory. This is relevant in a wealth of physical situations; specifically, we focus here on the control of nanodevices with stochastic dynamics. Therein, one is typically interested in minimizing the entropy production \cite{schmiedl_optimal_2007,gomez-marin_optimal_2008,aurell_optimal_2011,aurell_refined_2012,muratore-ginanneschi_extremals_2014,zhang_work_2020,loos2024universal} or the connection time \cite{hegerfeldt_driving_2013,hegerfeldt_high-speed_2014,plata_finite-time_2020,prados_optimizing_2021,patron_thermal_2022,patron_minimum_2024} between given initial and target states. In the lab, the controlled parameters are usually restricted to a certain finite region. For example, the stiffness $\kappa$ of a harmonic trap is generally limited to a nonnegative range $\kappa_- \le \kappa\le \kappa_+$~\footnote{Negative values of the stiffness are difficult to implement in the lab, although they have been realized employing feedback techniques~\cite{albay_realization_2020}.}, depending on the specific technique used to realize the harmonic potential. As a rule, these inequalities make variational calculus unreliable for solving the optimization problem: More sophisticated techniques from optimal control theory are necessary.

Despite recent advances in the control of colloidal particles~\cite{guery-odelin_driving_2023,blaber_optimal_2023}, important aspects remain challenging. Notably, most of the theoretical~\cite{schmiedl_optimal_2007,aurell_optimal_2011,aurell_refined_2012,martinez_engineered_2016,li_shortcuts_2017,plata_optimal_2019,plata_finite-time_2020,zhang_work_2020,plata_taming_2021,patron_thermal_2022,patron_minimum_2024,guery-odelin_driving_2023} and experimental results~\cite{ciliberto_experiments_2017,martinez_colloidal_2017} pertain to the overdamped regime, in which inertial effects are negligible. Therein, accelerating the connection between equilibrium states of a particle confined in an arbitrary potential has been shown to be feasible, by using several inverse engineering techniques~\cite{martinez_engineered_2016,plata_taming_2021,li_shortcuts_2017} imported from the field of shortcuts to adiabaticity~\cite{guery-odelin_shortcuts_2019}. Also, optimal protocols that minimize certain figures of merit have been investigated, e.g., minimization of irreversible work (entropy production) for a given connection time $t_f$, both for unbounded~\cite{schmiedl_optimal_2007,aurell_optimal_2011,aurell_refined_2012,zhang_work_2020} and bounded~\cite{plata_optimal_2019} stiffness. 

Extending control of colloidal particles to the underdamped regime is critical, since it is the most general description. Experimentally, underdamped dynamics accurately describes nanomechanical resonators~\cite{le_cunuder_fast_2016,Dago2022PRL} or levitated particles in moderate vacuum~\cite{Rondin2017NN,Rademacher2022PRL,raynal_shortcuts_2023}. Theoretically, the control problem is significantly more complex and more poorly understood than its overdamped counterpart. Inverse engineering techniques have been used but limitations appear~\cite{chupeau_engineered_2018}---and optimal control brings about new challenges. Optimal protocols, in general, involve non-conservative, velocity-dependent driving forces \cite{muratore-ginanneschi_extremals_2014,li_shortcuts_2017,li_geodesic_2022}.
If the force is restricted to be conservative, as done in this Letter, the optimal driving involves Dirac-delta type discontinuities~\cite{gomez-marin_optimal_2008}---which are not realizable in the lab, since they involve infinite values of the stiffness.

\begin{figure}
    \centering
    \includegraphics[width=3in]{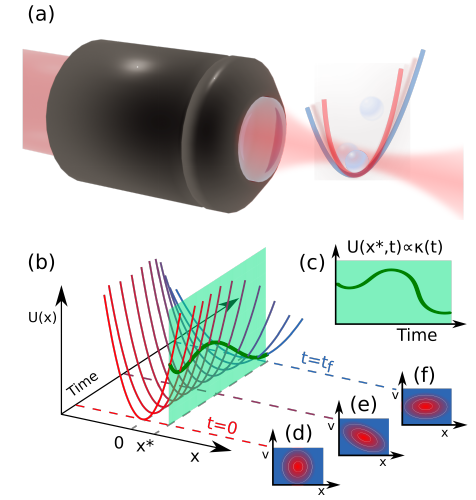}
    \caption{Principle of the experiment. (a) A nanoparticle is optically levitated by a highly focused laser. The harmonic confinement of the particle, i.e., trap stiffness, is tuned through the trapping laser power. (b) The trap stiffness $\kappa$ is changed in time according to a prescribed protocol (c), to bring the particle probability density function from an initial equilibrium state at time $t=0$ (d), to a final equilibrium state at time $t=t_f$ (f), passing through a series of out-of-equilibrium states (e). The green section cutting plot (b) at an arbitrary $x=x^*$ provides a direct visualization of the time dependence of $\kappa(t)$. The connection can be achieved in a plethora of ways, and we seek for optimal ones (timewise or energywise).}
    \label{fig:scheme}
\end{figure}

Our main goal is the optimal control of an underdamped particle, bringing to bear the practical bounds to achieve the optimal realizable protocol. In particular, we aim at minimizing the irreversible contribution to the work for a harmonically confined particle, with stiffness $\kappa$ that can be tuned at will in a certain finite interval $0\leq \kappa_- \leq \kappa \leq \kappa_+$. Both the initial and target states correspond to equilibrium, for the initial and target values of the stiffness.  It is worth stressing that the stiffness is the only parameter we control, i.e., the potential is harmonic, conservative, and confining at all times. Our theoretical results are tested by performing an experiment on a levitated nanoparticle trapped in an optical confining potential. The principle of the experiment, described below, is illustrated in Fig.~\ref{fig:scheme}.

\textit{Model.-} 
Consider a Brownian particle of mass $m$, position $\vect{x}$ and velocity $\vect{v}$, immersed in a fluid at constant temperature $T$. The collisions with the fluid molecules result in a damping force $-\gamma \vect{v}$, where $\gamma$ is a constant friction coefficient, plus a fluctuating force $\vect{\xi}(t)$ such that
$\av{\vect{\xi}(t)}=0$, $\av{\vect{\xi}(t)\vect{\xi}^T(t')}=2\gamma k_B T  I \delta(t-t')$,
where $I$ represents the identity matrix and $k_B$ is the Boltzmann constant. If the particle is trapped along a certain direction by the harmonic potential $U(x,t)=\kappa(t)x^2/2$, the projected motion obeys
$\dot{x}=v$, $m \dot{v}=-\kappa x-\gamma v+\xi$.
Equivalently, 
the probability density function $p(x,v,t)$ evolves through the Klein-Kramers equation~\cite{gardiner_stochastic_2009}
\begin{equation}
\label{eq:kkdyn}
\partial_{t} p=-\partial_{x}\cbr{v p}+\frac{1}{m} \partial_{v}\cbr{\kappa x p+\gamma v p+\frac{\gamma k_{B} T}{m} \partial_{v} p}\,,
\end{equation}
which admits the equilibrium solution for constant $\kappa$
\begin{equation}
\label{eq:eqdist}
p_{\text{eq}}(x,v|\kappa)=\frac{\sqrt{m \kappa}}{2 \pi k_B T}\exp\cbr{-\frac{mv^2+\kappa x^2}{2 k_B T}}\,.
\end{equation}
The above notation emphasizes that, in typical experimental setups, $\kappa$ is a controllable parameter while $T$ is fixed. For an initial Gaussian centered at $(x=0,v=0)$, as in equilibrium, Eq.~\eqref{eq:kkdyn} preserves this property throughout the evolution of $p(x,v,t)$. The system's state is then completely characterized by the vector $\vect{u}$ of its second moments\new{---see Sec.~I of~\cite{SM}.}
In terms of 
\begin{subequations}
\begin{align}\label{eq:w}
\w=&\frac{\gamma^2 \av{x^2}}{k_B T m}, &   
\y=&\frac{\gamma \av{xv}}{k_B T}, &
\z=\frac{m \av{v^2}}{k_B T}\,, \\
s=&{\gamma t}/{m} , & k=&{m\kappa}/{\gamma^2} ,
\end{align}
\end{subequations}
the evolution of the system's state reads
\begin{equation}
\label{eq:constr}
\dot{\uv} = \fv(\uv;k)\equiv M_k \uv +2 \vect{e}_3 ,
\end{equation}
in dimensionless variables, with $\dot{\uv}\equiv d\uv/ds$, and 
\begin{equation}
\label{eq:matrixmk}
    M_k=\begin{pmatrix}
        0 & 2 & 0 \\
        -k & -1 & 1 \\
        0 & -2k & -2 
    \end{pmatrix} ,
    \quad 
    \vect{e}_3=\begin{pmatrix}
    0 \\
    0 \\
    1 
\end{pmatrix}\,.
\end{equation}
We search for \new{swift state-to-state transformations (SSTs)}~\cite{guery-odelin_driving_2023} between an initial state $p_{\text{eq}}(x,v|\kappa_i)$
and a target equilibrium distribution  $p_{\text{eq}}(x,v|\kappa_f)$ corresponding to different confinements, or, equivalently, between
\begin{equation}
\label{eq:bcond}
\uv_i= \begin{pmatrix}
    1/k_i \\
    0 \\
    1 
\end{pmatrix}\,\quad \text{and}\quad
\uv_f= \begin{pmatrix}
    1/k_f \\
    0 \\
    1 
\end{pmatrix}\,.    
\end{equation}
To do so, we allow $k(s)$ to vary with arbitrary speed, and even to make discontinuous jumps, but require it to remain within prescribed bounds: 
\begin{equation}
\label{eq:range-k}
0\le
k_- \le k \le k_+ 
\,,
\end{equation}
with $k_-<k_{i,f}<k_+$. We ask (i) whether the connection can be completed in a given finite time $s_f$  by suitably steering $k$ within the prescribed range~\eqref{eq:range-k}, without waiting for the system's spontaneous relaxation; (ii) 
\new{what the SST protocol, from the plethora of them making the connection, minimizes the associated average external work.}
Since the cumulative average work at time $s$ is 
\begin{equation}
\label{eq:work}
 {W(s)}=k_B T \int_0^{s}ds'\,\frac{\dot{k}(s')}{2}  u_1(s')\, ,
\end{equation}
we then look for the protocol $k(s)$ that minimizes ${W(s_f)}$.

The main tool for deriving our analytical results is Pontryagin's maximum principle (PMP) for minimizing a prescribed cost $\mathcal{C}=\int_0^{s_f} ds \, \mathcal{L}(\uv;k) $, functional of the whole trajectory~\cite{pontryagin_mathematical_1987,liberzon_calculus_2012}. 
In a nutshell, 
PMP states that the time-dependent protocol $k(s)$ that minimizes $\mathcal{C}$ in turn maximizes the Hamiltonian $\mathcal{H}(\uv,\pv;k)=\pzero\mathcal{L}(\uv;k)+\pv \cdot \fv$, where $\pzero \leq  0$ is a constant and $\pv$ is a vector of conjugated momenta, with $\uv$ and $\pv$ obeying the canonical equations $\dot{\uv}=\nabla_{\pv}\mathcal{H}=\fv$ and $\dot{\pv}=-\nabla_{\uv}\mathcal{H}$, with $(\pzero,\pv^T)\neq (0,0,0,0)$. 
\new{Note that PMP provides a necessary, but in general not sufficient, condition for optimality. See Sec.~II of~\cite{SM} for more details.} 

\textit{Minimal time.-} To address our first question, we search for the minimum connection time $s_{\text{th}}$ between $\uv_i$ and $\uv_f$. Here, $\mathcal{L}=1$ and Pontryagin's Hamiltonian thus reads 
$\mathcal{H}=\pzero+\pv \cdot \fv(\uv;k)$, 
whose maximum can only be located at the extreme values $k=k_{\pm}$~\footnote{Imposing $\pdv{H}{k}=0$ does not give a solution, see Sec.~III of~\cite{SM} for details.}. Therefore, the minimum time is provided by a protocol alternating at least three branches with constant $k=k_{\pm}$, known as ``bang-bang'' in control theory~\cite{chen_fast_2010,hegerfeldt_driving_2013,hegerfeldt_high-speed_2014,lu_fast_2014,ding_smooth_2020,prados_optimizing_2021,ruiz-pino_optimal_2022,patron_thermal_2022}. The switching times and the total time $s_{\text{th}}$ \new{are numerically computed in Sec.~III of~\cite{SM}.} If a protocol $k^{\star}(s)$ evolves $\uv_i$ into $\uv_f$ in a time $s_{\text{th}}$, the same connection can be made in any time $s_f > s_{\text{th}}$~\footnote{E.g., by getting to $\uv_f$ in a time $s_{\text{th}}$ by applying $k^{\star}(s)$, and then keeping the system at equilibrium for a time $s_f-s_{\text{th}}$ by setting $k(s)=k_f$.}. Therefore, our first result is that $s_f \ge s_{\text{th}}$ is a sufficient condition to realize the desired connection. 
\begin{figure}
    \centering    
    \includegraphics[width=3in]{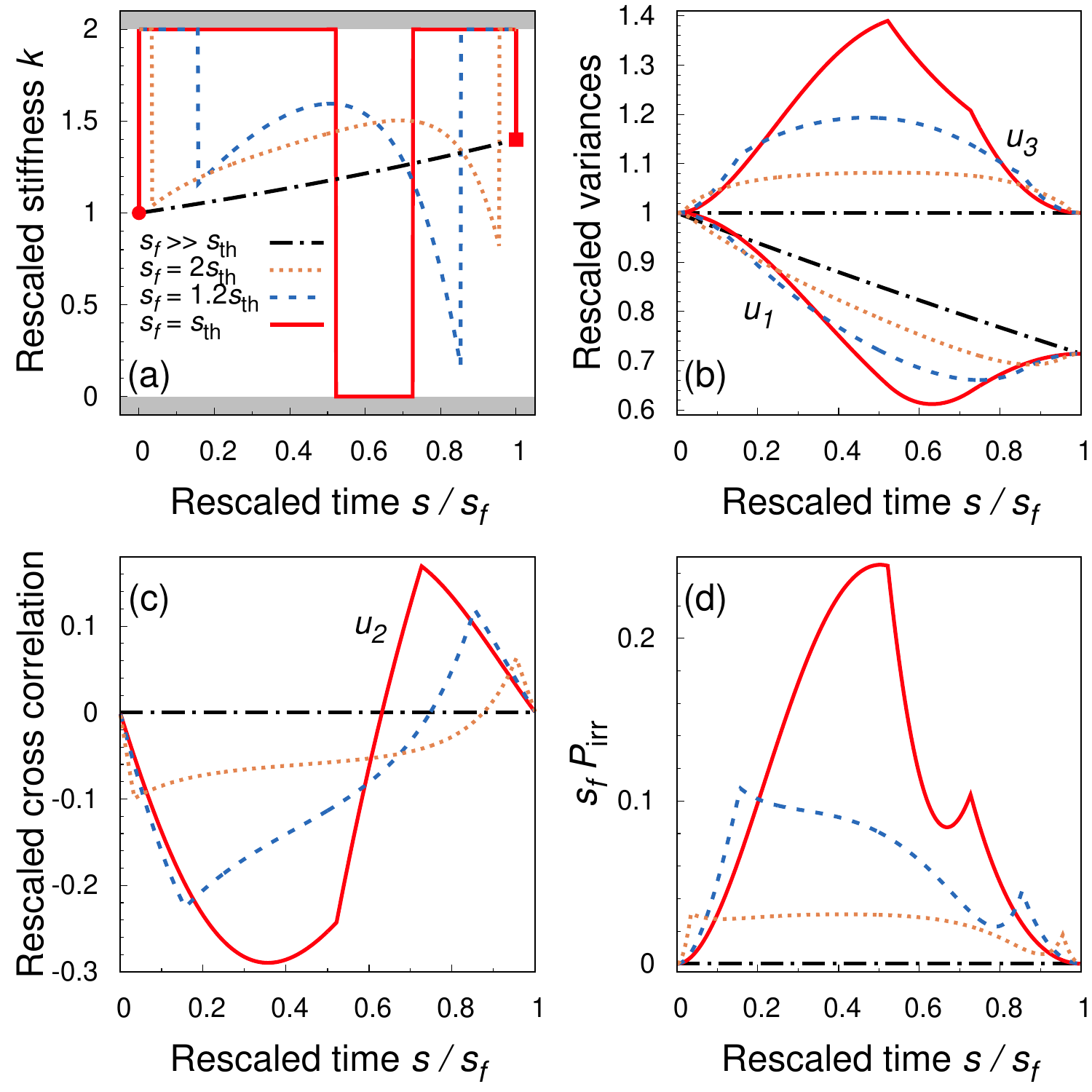}
    \caption{Minimal work protocols. Panel (a) shows 
    \new{the optimal protocols for different values of the rescaled connection time $s_f$.} The dash-dotted black curve corresponds to the quasi-static limit,  \new{$s_f\gg s_{\text{th}}$, while the solid red curve corresponds to the minimal-time protocol, $s_f=s_{\text{th}}$.} Shaded areas denote forbidden values of the stiffness. In panels (b) and (c), the corresponding evolution 
    of the rescaled moments $u_1$, $u_2$ and $u_3$ is shown.
    Finally, panel (d) exhibits the time dependence of the irreversible part of the average power (the integrand of Eq.~\eqref{eq:wirr}). Plot axes are rescaled in such a way that the total area below each curve is proportional to the irreversible work done during each process, which decreases with $t_f$ as expected, 
    Here, $k_i=1$, $k_f=1.4$, $k_-=0$, $k_+=2$.}
    \label{fig:minW}
\end{figure}

\textit{Minimal work.-} Our next step is to search for the protocol that minimizes the total work, among those achieving the connection in a time $s_f \ge s_{\text{th}}$. The total average work~\eqref{eq:work} splits into two contributions: (i) a reversible one, which is physically related to the free energy variation; and (ii) an irreversible one, which is a functional of $\uv$ and $k$, vanishing in the quasistatic limit (see Sec.~IV of~\cite{SM}):
\begin{equation}
    \label{eq:wirr}
{W_{\text{irr}}}=k_B T \int_0^{s_f}ds\,\underbrace{\cbr{\frac{\w}{\w\z-\y^2}+\z-2}}_{P_{\text{irr}}}\,.
\end{equation}
The optimal solution is found by minimizing either ${W_{\text{irr}}}$ directly or the excess work with respect to the energy variation $\mathcal{C}=\int_0^{s_f}ds\,\cbr{\z-1}$~\footnote{The two approaches are equivalent, because the boundary conditions are fixed: we will pursue the latter, which is slightly easier, see Sec.~IV of~\cite{SM}.}.
Now $\mathcal{L}=u_3-1$ and, choosing $\pzero=-1$,
$\mathcal{H}=-\z+1+\pv \cdot \fv(\uv;k)$.
The maximum of $\mathcal{H}$ as a function of $k$ can be at either a value $k_0$ such that $\left.\partial\mathcal{H}/\partial k\right|_{k=k_0}=0$ or the bounds $k_\pm$. The first option yields, as shown in Sec.~IV of~\cite{SM},
\begin{equation}
\label{eq:ksol}
    \dot{k}_0=3k_0+\frac{6}{\w}+3\frac{\y}{\w}-9\frac{\z}{\w}+10\frac{\y^2}{\w^2}-8\frac{\y \z}{\w^2}+8\frac{\y^3}{\w^3}\,,
\end{equation}
which, if assumed to be valid $\forall s\in(0,s_f)$, leads to a protocol $k_0(s)$ that violates the bounds. In fact, this $k_0(s)$ coincides with the variational solution, which has Dirac-delta discontinuities at the initial and final times~\footnote{Similarly to the situation found in Ref.~\cite{gomez-marin_optimal_2008}, although the boundary conditions are different---see Sec.~IV of~\cite{SM} for details}, and thus can only be valid for unbounded stiffness. Inspired by the structure of the unbounded solution, we look for a three-stage protocol in the bounded case:
\begin{equation}\label{eq:optimal-bang-EL-bang}
k(s) =
\begin{cases} 
    k_\pm, & s \in [0, s_1],  \quad\;\text{(bang)}, \\
    k_0(s), & s \in [s_1, s_2],  \quad\text{(Euler-Lagrange)}, \\
    k_\pm, & s \in [s_2, s_f], \quad\text{(bang)}.
\end{cases}
\end{equation}
This three-stage optimal protocol, together with the corresponding time evolution of the correlations and the irreversible output power, is illustrated in Fig.~\ref{fig:minW}.

Note that we expect a time-scale separation between the (slow) control dynamics and the (fast) system thermalization for $s_f \to \infty$. Then, the optimal solution tends to the overdamped limit one~\cite{muratore-ginanneschi_extremals_2014,muratore2014, sanders2024optimal}, namely
\begin{equation}
   k(s)=\left(\frac{s_f-s}{\sqrt{k_i} s_f}+\frac{s}{\sqrt{k_f} s_f}\right)^{-2}\,,
   \label{eq:adia}
\end{equation}
which can be realized without exceeding the range $k \in [k_i,k_f] \subset [k_-,k_+]$. We thus expect the optimal solution in the time interval $[s_1,s_2]$, determined by $k_0(s)$, to verify $k_-\leq k_0(s) \leq k_+$ if $s_f \gg s_\text{th}$. A three-stage optimal solution of the  ``bang--Euler-Lagrange--bang'' type, as proposed above, should thus always be possible under these conditions.  If $s_f \gtrsim s_\text{th}$, a more complex shape may be required: see Sec.~V of~\cite{SM} for an explicit example.

\textit{Experimental results.-} 
To experimentally test the derived optimal protocol, we use an optically levitated particle in moderate vacuum and at room temperature ($T_0=290$~K)~\cite{raynal_shortcuts_2023}. 
An infrared laser is strongly focused by an objective to generate an optical tweezer trapping a 68~nm-radius spherical silica particle in a harmonic potential of natural frequency $\Omega\approx 2\pi \times 100$~kHz. The trap stiffness $\kappa=m \Omega^2$, with $m\approx 2.4 \times 10^{-18}$~kg the particle mass, is directly proportional to the trapping laser power $P_\text{las}$.  The system damping is enforced via the gas pressure $p_\text{gas}$ in the experiment chamber. Specifically, we use $p_\text{gas} \approx 5$~hPa, corresponding to  $\tau^{-1}=\gamma/m=2\pi \times 4.4$~kHz $ \ll \!\Omega$, which belongs in the underdamped regime\new{---note that $\tau$ is the natural relaxation time.} Using an ancillary laser, we record the particle dynamics $x(t)$. 

\begin{figure}
\centering
\includegraphics[width=3in]{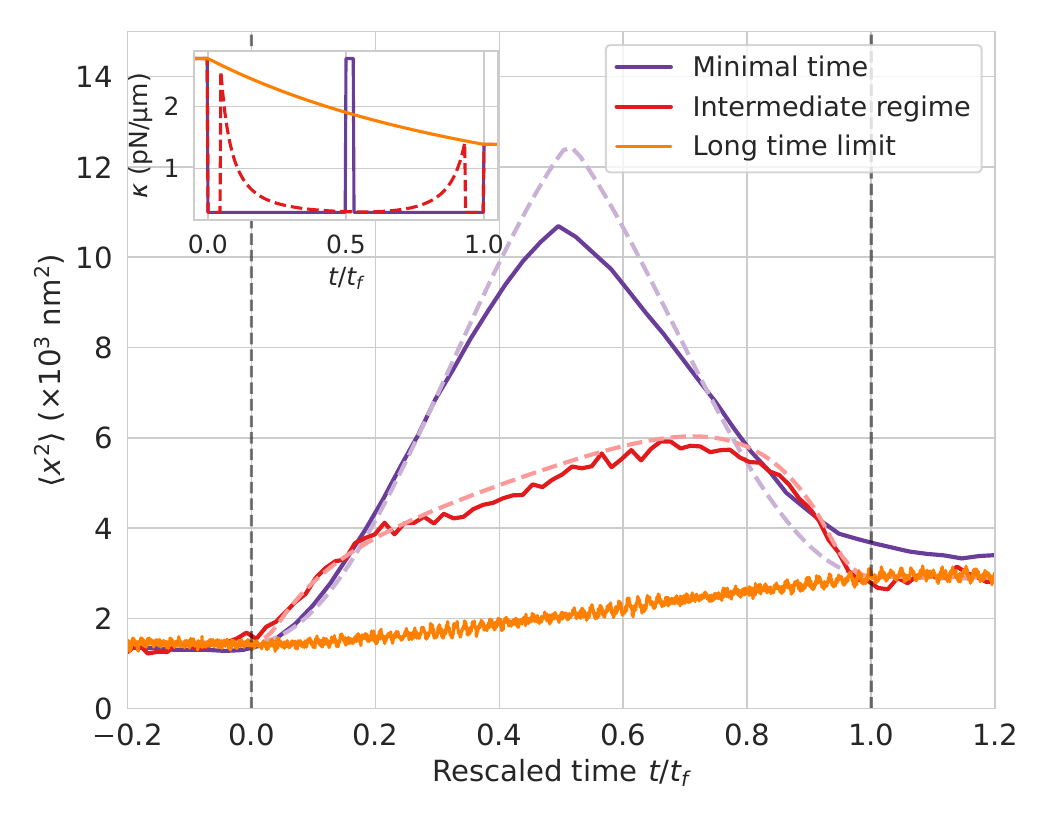}
\caption{\new{
Position variance for the experimental realization of optimal protocols. The plotted data correspond to 30\,000 realizations of the minimal time (violet, $t_f=t_\mathrm{th}=7~\mu$s), bounded optimal in the intermediate time regime  (red, $t_f=1.8\ t_\mathrm{th}=12.6~\mu$s), and long-time limit (orange, $t_f=480~$µs$\gg t_\mathrm{th}$) protocols. All experiments correspond to the same bounds and compression factor of  $\chi=0.5$.
The experimental variance evolutions fairly agree with analytical expectations (dashed lines). 
Vertical black dashed lines mark the initial and final times of the protocols, which are shown in the inset.
}}\label{fig:prot_class}
\end{figure}
As discussed earlier, a critical limitation of experimental realizations of thermodynamic protocols is that the control parameters are limited to a finite range.  We control the trap stiffness through the trapping laser power using an acousto-optic modulator, 
which technically limits the ratio of available stiffnesses $\kappa_+/\kappa_-$ to a factor of typically ten. Specifically, we have $\kappa_-=0.29$~pN/µm ($\Omega_- =2\pi \times 55$~kHz) and $\kappa_+\approx 2.77$~pN/µm ($\Omega_+ =2\pi \times 170$~kHz). We address decompression protocols with compression factor $\chi=\kappa_f/\kappa_i=0.5$, choosing $\kappa_i=\kappa_+$.

\new{Under these conditions, the minimal threshold time for a protocol is $t_\mathrm{th} \approx 7~$µs. We experimentally realize the derived SST optimal protocols for the minimization of the irreversible work~\cite{supportingdata}. They are shown in the inset of Fig.~\ref{fig:prot_class}: the minimal time protocol (violet) with a final time $t_f = t_\mathrm{th}$; an intermediate time optimal protocol (red, $t_f = 1.8\ t_\mathrm{th}$); and a long-time-limit protocol following Eq.~(\ref{eq:adia}) (orange, $t_f = 480~\text{µs} \gg t_\mathrm{th}$).}
As shown in Fig.~\ref{fig:prot_class}, the resulting position variance $\langle x^2(t)\rangle$ reaches its 
final equilibrium value close to $t_f$, matching the theoretical expectation. 
\new{Thus, the bounded optimal protocols successfully drive the system to equilibrium at the targeted final time. For the minimum time protocol, a small discrepancy between the experimental and theoretical processes is observed, which we attribute to the challenge of precisely matching experimental parameters: for $t_f=t_{\mathrm{th}}$, the duration of the bangs is close to the typical response time of the experimental setup.}

\begin{figure}
\centering
\includegraphics[width=3in]{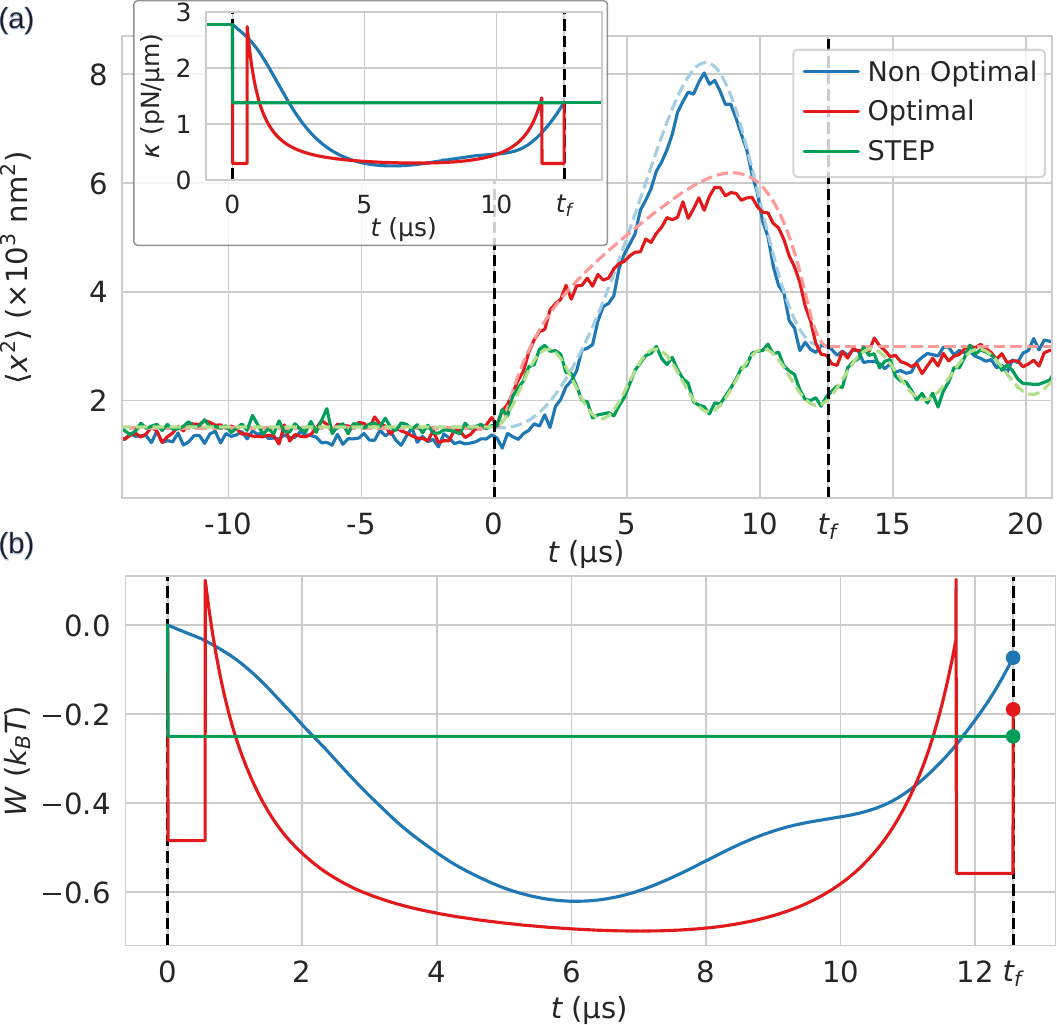}
\caption{(a) Position variance of the particles. The plotted data correspond to 30\,000 realizations of STEP (green), non-optimal SST (blue), and bounded optimal SST  (red) protocols, whose associated stiffnesses are shown in the inset.  The final time of both the non-optimal SST and the bounded optimal SST is $t_f\approx 12.6$~µs. A fit of the STEP relaxation provides access to experimental parameters (red dashed line). The experimental data agree with analytical expectations (dashed lines). Vertical black dashed lines mark the starting and final times of the SST protocols. The optimal protocol is identical to the intermediate time regime protocol in Fig.~\ref{fig:prot_class}, and is reproduced for clarity.
(b) Experimental cumulative work performed for the STEP (green), non-optimal SST (blue), and bounded optimal SST (red) protocols.
}
\label{fig:exp}
\end{figure}

\new{For $t_f=t_{\mathrm{th}}$, there is only one protocol, the brachistochrone, that makes the connection, at an expensive work cost; for $t_f\gg t_{\mathrm{th}}$, any regular protocol would be quasistatic and give a very small irreversible work, virtually indistinguishable from that for the optimal one---given by Eq.~\eqref{eq:adia}. From the experimental point of view, intermediate times, $t_f>t_{\mathrm{th}}$ with $t_f/t_{\mathrm{th}}=O(1)$, is thus the most interesting regime for our protocols with minimal work because it offers an advantageous trade-off between energetic cost and protocol duration. Therefore, we have considered two SST protocols with the same finite connection time $t_f=1.8\ t_{\mathrm{th}}$, shown in the inset of Fig.~\ref{fig:exp}: the optimal one, given by Eq.~\eqref{eq:optimal-bang-EL-bang}, and a non-optimal one. The latter relies on choosing the lowest-order polynomial that makes the connection~\cite{chupeau_engineered_2018,raynal_shortcuts_2023}. For reference, we also consider a STEP protocol, corresponding to a sudden change of stiffness from $\kappa_i$ to $\kappa_f$.}

As previously, we looked into the time evolution of the particle position variance driven by these protocols, bounded optimal, non-optimal SST, and STEP, which is shown in Fig.~\ref{fig:exp}\new{-(a)}. As expected, in the case of the STEP protocol, the variance presents oscillations with an exponential decay of characteristic time $\tau=m/\gamma\approx 3.8 \times 10^{-5}$~s, and it takes infinite time to exactly reach the target equilibrium state. In contrast, both the SST and the optimal protocol allow one to reach the target state at the desired final time \new{$t_f=\tau/3=1.8\ t_{\mathrm{th}}$.} 

We then compute the average experimental work required to drive the particle according to the three protocols, using the definition~\eqref{eq:work}---as detailed in Secs.~VI and VII of~\cite{SM}. The cumulative work resulting from the experimental measurements is shown in Fig.~\ref{fig:exp}-(b). At the final time $t_f$, we confirm that the bounded optimal protocol allows us to reach equilibrium with average work $W_\text{optimal}=-0.19~k_B T$, lower than for the non-optimal SST protocol under similar conditions, $W_\text{SST}=-0.07~k_B T$. The STEP protocol provides an even lower work up to $t_f$, $W_\text{STEP}=-k_B T/4$, but the system has not reached the target state.

\textit{Conclusions.-} The results shown in Figs.~\ref{fig:exp} account for a laboratory realization of the minimal-work transition between two equilibrium states of a levitated particle subject to thermal noise and viscous friction, in a regime where inertial effects are relevant. The control protocol for the confining potential is applied within an actual experimental setup, showing that it can be implemented up to a very good degree of approximation. The target state is reached with the same accuracy as with the non-optimized SST protocol studied in~\cite{raynal_shortcuts_2023}, but with a significantly lower energetic cost. The naive STEP protocol obtained by abruptly changing the confining stiffness and waiting for relaxation leads instead to a state which is far from being thermalized at the prescribed final time. 

The present experiment puts forward a proof of principle for optimal control protocols at the nanoscale, by showing that fast optimized transitions are realizable in the lab with levitated particles. Moreover, the results reported here are expected to be relevant, for instance, for the realization of thermal engines at scales where thermal effects are not negligible and stochastic thermodynamics must be taken into account~\cite{Dechant2015PRL,liRealizationAllopticalUnderdamped2024,messageExtremetemperature2025,chatterjee_optimal_2025}. Minimizing work done on the system---i.e., maximizing work extracted from it---at constant bath temperature is a typical building block in the design of efficient thermodynamic cycles: e.g. this corresponds to the optimal isothermal branches of maximum-power irreversible heat engines~\cite{plata_building_2020,prieto-rodriguez_maximum-power_2025}.

From the theoretical perspective, a challenging point to be further explored is the case of connection times close to the minimum time $s_{\text{th}}$. Therein, minimal work is usually not given by the three-stage process considered here (bang--Euler-Lagrange--bang) and more complex behaviors need to be taken into account. \new{Also, we recall that PMP provides a necessary condition for optimality. Rigorously proving there is not any bounded SST protocol providing an even lower value for the considered cost functions, either connection time or irreversible work, is an open mathematical problem---worth of being investigated but beyond the scope of this Letter.} Another research line concerns the more general case of nonharmonic confinements, which has recently been theoretically discussed~\cite{sanders2024optimal,sanders2024minimalworkprotocolsinertialparticles}---but in the limit of small inertial effects with unbounded stiffness. Finally, an interesting field of investigation involves the extension of control techniques to active-matter systems~\cite{Shankar_2022,massana2022rectification}, for which some theoretical results are already available~\cite{baldovin2023,davis2023active}.

\begin{acknowledgments}
MB thankfully acknowledges useful discussions with J. Sanders and P. Muratore-Ginanneschi. MB was supported by ERC Advanced Grant RG.BIO (Contract No.~785932). CAP and AP acknowledge financial support from Grant PID2024-155268NB-I00 funded by MICIU/AEI/10.13039/501100011033/ FEDER, UE, and also from the applied research and innovation Project PPIT2024-31833, cofunded by EU--Ministerio de Hacienda y Función Pública--Fondos Europeos--Junta de Andalucía--Consejería de Universidad, Investigación e Innovación. CAP, AP, and ET acknowledge financial support from Grant ProyExcel\_00796 funded by Junta de Andalucía's PAIDI 2020 program. CAP acknowledges the funding received from European Union’s Horizon Europe–Marie Skłodowska-Curie 2021 program through the Postdoctoral Fellowship with Reference 101065902 (ORION). This work is supported by the ANR projects OPLA (ANR-20-CE30-0014) and FENNEC (ANR-23-CE30-0042).
\end{acknowledgments}


\bibliography{biblio,Mi-biblioteca-24-abr-2025}

\end{document}


\title[Supplementary Material for ``Optimal control of levitated nanoparticles through finite-stiffness confinement'']{Supplementary Information for ``Optimal control of levitated nanoparticles through finite-stiffness confinement''}

\author{Marco Baldovin}
 \email{marco.baldovin@cnr.it}
\affiliation{Institute for Complex Systems, CNR, Universit\`a Sapienza, I-00185, Rome, Italy}
\affiliation{LPTMS, CNRS, Universit\'e Paris-Saclay, F-91405, Orsay, France}
\author{Ines Ben Yedder}
\affiliation{CentraleSup\'elec, LuMIn, Université Paris-Saclay, ENS Paris-Saclay, CNRS, F-91405, Orsay, France}
\author{Carlos A. Plata}
\affiliation{F\'{\i}sica  Te\'orica, Multidisciplinary Unit for Energy Science, Universidad  de  Sevilla, Apartado de Correos 1065, E-41080, Sevilla, Spain}
\author{Damien Raynal}
\affiliation{CentraleSup\'elec, LuMIn, Université Paris-Saclay, ENS Paris-Saclay, CNRS, F-91405, Orsay, France}
\affiliation{Centre of Light for Life (CLL) and Institute for Photonics and Advanced Sensing (IPAS), The University of Adelaide, 5005, Adelaide, SA, Australia}
\author{Lo\"ic Rondin}
\affiliation{CentraleSup\'elec, LuMIn, Université Paris-Saclay, ENS Paris-Saclay, CNRS, F-91405, Orsay, France}
\author{Emmanuel Trizac}
\affiliation{\'Ecole  Normale  Sup\'erieure de Lyon, F-69342, Lyon, France}
\author{Antonio Prados}
\affiliation{F\'{\i}sica  Te\'orica, Multidisciplinary Unit for Energy Science, Universidad  de  Sevilla, Apartado de Correos 1065, E-41080, Sevilla, Spain}

\maketitle

We provide here supplementary material to support our findings.  \new{In Section~\ref{sec:evol-eqs}, we present a brief derivation of the evolution equations for the relevant moments from the Fokker-Planck equation.} In Section~\ref{sec:pontryagin} we recall the basic concepts of Pontryagin's theory, and we apply it in Section~\ref{sec:minimaltime} to compute the minimal time protocols for the dynamics studied in the main text. In Section~\ref{sec:min-work} we provide explicit derivations for (i) the form of the cost function to minimize irreversible work and (ii) the structure of the solution for the optimal protocol. An example for a connection time close to the minimal time, where the three-stage solution in the main text (bang--Euler-Lagrange piece--bang) is thus not valid, is explicitly discussed in Section~\ref{sec:shorttimes}. Sections~\ref{sec:calibration} and~\ref{sec:computation} are devoted to the discussion of experimental aspects: in the former, we provide details on the calibration procedure; in the latter, we explain the procedure we use to compute the work form experimental data.

\new{
\section{Derivation of the dynamical equations for the moments}\label{sec:evol-eqs}}

\new{In this Section we provide a synthetic derivation of Eqs.~(4)-(5) of the main text, describing the dynamics of the moments. Our starting point is the Klein-Kramers equation
\begin{equation}
\label{eq:kkdyn}
\partial_{t} p=-\partial_{x}\cbr{v p}+\frac{1}{m} \partial_{v}\cbr{\kappa x p+\gamma v p+\frac{\gamma k_{B} T}{m} \partial_{v} p}\,,    
\end{equation}
which is an equivalent description for the Langevin evolution of a Browninan particle of mass $m$, subject to an external potential $U(x)=\kappa x^2/2$, with damping coefficient $\gamma$ and temperature $T$. Here $p\equiv p(x,v,t)$ represents the probability that, at time $t$, the Brownian particle is found at position $x$ with velocity $v$. We notice that dividing both members by $p$, we can rewrite  Eq.~\eqref{eq:kkdyn} in the form
\begin{equation}
\label{eq:kkdynlog}
    \partial_t \ln p = -v \partial_x \ln p + \frac{\kappa}{m}x \partial_v \ln p +\frac{\gamma}{m}+\frac{\gamma}{m}v\ln p + \frac{\gamma k_B T}{m^2}\partial_v^2 \ln p + \frac{\gamma k_B T}{m^2} (\partial_v \ln p)^2\,,
\end{equation}
which will reveal quite convenient in a moment.}

\new{Since in this work we limit ourselves to cases in which the boundary conditions at $t=0$ and $t=t_f$ are equilibrium distributions, and the dynamics is linear, the form of the solution must be Gaussian at all times. 
From Eq.~\eqref{eq:kkdyn} one finds that $ d\av{x}/dt=\av{v}$ and $ d\av{v}/dt=-\frac{\kappa}{m}\av{x}-\frac{\gamma}{m}\av{v}$.
As a consequence of our choice of the potential (centered at zero), we have therefore that the equilibrium distribution must be centered at zero for all times, meaning that $\av{x}$ and $\av{v}$ vanish.  These considerations allow us to write the solution in terms of the covariances only: in its most general form, it reads
\begin{equation}
p(x, v, t)=\frac{1}{2 \pi (\av{x^2}\av{v^2}-\av{xv}^2)}\exp\cbr{-\frac{1}{2}\frac{\av{v^2}x^2-2\av{xv}xv+\av{x^2}v^2}{\av{x^2}\av{v^2}-\av{xv}^2}}\,.
\end{equation}}
\new{
By plugging the above Gaussian solution into Eq.~\eqref{eq:kkdynlog}, and isolating terms proportional to $x^2$, $xv$ and $v^2$, we obtain a system of ordinary differential equations for the moments, namely
\begin{subequations}
\label{eq:constr}
\begin{eqnarray}
&\dot{u}_1=&2u_2 \\   
&\dot{u}_3 =&-2ku_2-2u_3+2 \\ 
&\dot{u}_2=&u_3-ku_1-u_2\,,
\end{eqnarray}   
\end{subequations}
where we have introduced the dimensionless variables
\begin{equation}\label{eq:w}
\w=\frac{\gamma^2 \av{x^2}}{k_B T m}, \quad \quad 
\y=\frac{\gamma \av{xv}}{k_B T}, \quad \quad
\z=\frac{m \av{v^2}}{k_B T}\,, \quad \quad
s={\gamma t}/{m} , \quad \quad k={m\kappa}/{\gamma^2} .
\end{equation}
It is immediate to verify that Eq.~\eqref{eq:constr} coincides with Eqs.~(4) and~(5) of the main text.}
\vspace{2ex}

\section{Pontryagin's principle}
\label{sec:pontryagin}

The aim of Pontryagin's theory is to provide necessary conditions for the protocol minimizing a cost functional of the form
\begin{equation}
    \mathcal{C}[\vect{u};k]=\int_0^{t_f}dt\, \mathcal{L}(\vect{u}(t);k(t)),
\end{equation}
where $\mathcal{L}$ is a ``Lagrangian'', along the evolution of a generic dynamical system
\begin{equation}\label{eq:dynamics-u}
 \dot{\uv}(t)=\vect{f}[\uv(t);k(t)]\,   
\end{equation}
in the time interval $0\le t\le t_f$.

The first step is to define the Hamiltonian
\begin{equation}
\mathcal{H}(\uv,\pv;k)=\pzero\, \mathcal{L}(\uv;k)+\pv \cdot \vect{f}(\uv;k)\,,    
\end{equation}
where $\pv$ is the vector of the conjugated momenta of $\vect{u}$, while $\pzero \le 0$ is a constant. The evolution of the system in this enlarged symplectic space is given by the canonical equations
\begin{equation}
\label{eq:symdyn}
\dot{\uv}=\nabla_{\pv}\mathcal{H}\,,  \quad\quad \dot{\pv}=-\nabla_{\uv}\mathcal{H}\,.    \end{equation}
The minimization is then obtained by searching, at each time, for the value of $k$ that maximizes $\mathcal{H}$:
\begin{equation}
k(t)=\arg \sup_k \mathcal{H}(\uv,\pv;k)\,.
\end{equation}
This optimal $k(\uv,\pv)$, when inserted into the canonical equations, leads to an optimal trajectory $(\uv(t),\pv(t))$ in phase space. The candidate thus obtained, $\{k(t),\uv(t),\pv(t)\}$, must verify $(\varphi_0,\pv)\ne (0,\bm{0})$ for all $t\in [0,t_f]$ to provide a solution for the minimization problem. Over the solution 
$\{k(t),\uv(t),\pv(t)\}$ found in this way, the Hamiltonian is constant, independent of time. 

We note that, when no bounds are present on $k$, the above strategy provides the maximum of $\mathcal{H}$ from the condition $\partial\mathcal{H}/\partial k=0$, which leads to the usual Euler-Lagrange equations of variational calculus. In the variational calculus framework, $\pv$ plays the role of the Lagrange multipliers ensuring that the dynamical system defined by~\eqref{eq:dynamics-u} holds.


\new{Some comments on Pontryagin's approach are in order:
\begin{enumerate}
\item In some mathematical textbooks~\cite{liberzon_calculus_2012}, $\varphi_0$ is termed the ``abnormal'' multiplier. If $\varphi_0\ne 0$, it can be normalized to $\varphi_0=-1$ without loss of generality, within the framework of Pontryagin's \textit{maximum} principle. Here, we emphasize the word \textit{maximum} because one may alternatively employ an equivalent Pontryagin's \textit{minimum} principle, in which the Hamiltonian is minimized and the sign of $\varphi_0$ is reversed, normalized to $\varphi_0=+1$ when different from zero. We prefer to employ the maximum version of the principle because it better fits our physical intuition: with $\varphi_0=-1$, Pontryagin's Hamiltonian coincides with the usual Hamiltonian in classical mechanics problems. In some ``degenerate'' case, $\varphi_0=0$: this is the reason why we have kept it in our presentation---although in our problem we have found that $\varphi_0\ne 0$. For a geometrical intuition of the degenerate case, see e.g. the discussion on the maximum principle in Section 4.3 of Liberzon's book~\cite{liberzon_calculus_2012}.
\item 
Pontryagin's maximum principle provides a necessary condition for optimality, which in general is not sufficient. Therefore, in a general optimal control problem, one may be able to find different ``candidate'' solutions that fulfill Pontryagin's principle. In that case---which we have not found in our problem either, the candidate solution that gives the minimum cost is the only actual candidate, the others can be  disregarded.
\item 
For some linear problems, which are both linear in the dynamic variables and the controls, there are mathematical theorems that assure sufficiency for Pontryagin's approach. Nevertheless, they are not applicable to our case, since our system belongs to the class of bilinear problems---due to the terms proportional to $k u_i$ in the evolution equations. Therefore, proving that our controls actually give the minimum cost---either connection time or irreversible work---is a tough mathematical problem, beyond our goals.
\item 
Finally, we would like to stress that, as presented here, Pontryagin's approach is able to deal with bounds in the control function $k$ but not in the state variables $\uv$. If the latter are present, there are extensions of Pontryagin's principle to deal with this  situation, by including a new Lagrange multiplier in the mathematical description---for example, see chapter VI of Pontryagin's book~\cite{pontryagin_mathematical_1987}. In principle, one could think that this is the case we have in our physical problem, since both $u_1$ and $u_3$ verify the inequalities $u_1\ge 0$ and $u_3\ge 0$. Yet, it must be noted that one cannot reach a state with negative values of either $u_1$ or $u_3$ starting from a initial state in which they are not: the dynamics of the system prevent the variances from taking ``unphysical'' values. This is the reason why we have been able to use the version of Pontryagin's principle in which only the bounds in the control are considered.
\end{enumerate}
}
\begin{figure}
    \centering
    \includegraphics[width=.7\linewidth]{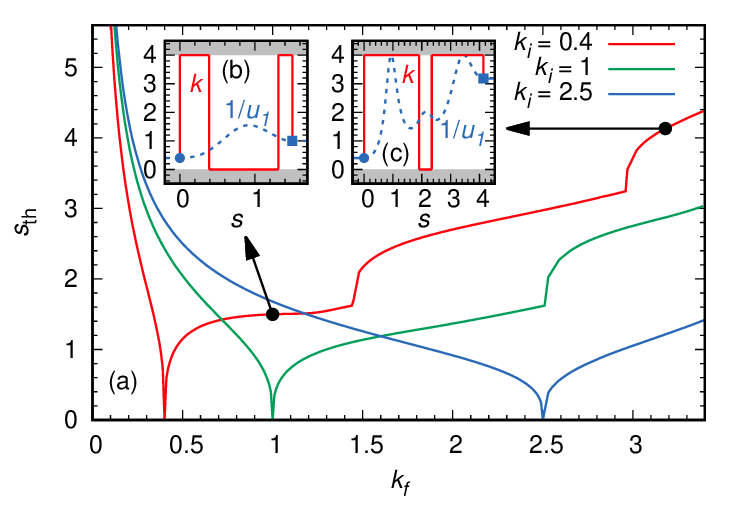}
    \caption{Minimal time protocols. In the main panel (a), the dependence of the minimal connection time for a three-step protocol~(\ref{eq:opt}) is shown as a function of the final (rescaled) stiffness $k_f$, for different values of $k_i$. Insets (b) and (c) show the explicit protocol $k$ (red, solid) and the inverse of the rescaled position variance, $1/u_1$ (blue, dashed), for the two cases ($k_i=0.4$, $k_f=1$) and ($k_i=0.4$, $k_f=3.18$). Shaded areas correspond to forbidden values of the stiffness. Here, $k_-=0$ and $k_+=4$.}
    \label{fig:mintime}
\end{figure}

\section{Minimal time}
\label{sec:minimaltime}

In terms of Pontryagin's principle, we need to minimize the cost
\begin{equation}
    \mathcal{C}\equiv s_{f}=\int_0^{s_f} ds \,,
\end{equation}
leading to the Hamiltonian
\begin{equation}
    \mathcal{H}(\uv,\pv;k)=\pzero+\pv \cdot \cbr{M_k \uv + 2 \vect{e}_3}\,.
\end{equation}
We need to find, at each time, the value of $k$ that maximizes $\mathcal{H}$. There are two possibilities: either $k$ is a solution of
\begin{equation}
\label{eq:dhzero}
\frac{\partial \mathcal{H}}{\partial k}=0\,,    
\end{equation}
or it coincides with one of the boundary values $k_{\pm}$. In the former case, condition~\eqref{eq:dhzero} leads to the identity
\begin{equation}
\label{eq:htime1}
     \w \py+ 2  \y \pz=0\,.
\end{equation}
Time-derivative yields, bearing in mind~\eqref{eq:symdyn},
\begin{equation}
\py \frac{\partial \mathcal{H}}{\partial \w}-\w \frac{\partial \mathcal{H}}{\partial \py}+2 \pz \frac{\partial \mathcal{H}}{\partial \y} -2 \y \frac{\partial \mathcal{H}}{\partial \pz} =0.
\end{equation}
Hence, making use of~\eqref{eq:dhzero},
\begin{equation}
\label{eq:htime2}
 \w \pw- \z \pz =0\,.   
\end{equation}
Taking again time derivative, one obtains
\begin{equation}
\label{eq:htime3}
2  \y \pw+ \z \py - 2 \pz =0 .
\end{equation}
Equations~\eqref{eq:htime1}--\eqref{eq:htime3} form a linear algebraic system for $\pv$, with determinant $2\w^2\ne 0$, which thus only admits the trivial solution
\begin{equation}
\pw=\py=\pz=0\,,
\end{equation}
In addition, for the time minimization problem, the Hamiltonian identically vanishes over the optimal trajectory~\cite{pontryagin_mathematical_1987,liberzon_calculus_2012}. This entails that $\varphi_0=0$, and therefore this candidate cannot be a solution of the minimization problem.  The above discussion implies that $\partial H/\partial k$ cannot vanish
over a finite time interval of the solution of the minimization problem. Therefore, depending on the sign of $\partial H/\partial k$, the optimal value of $k$ would be either $k_+$ or $k_-$. 
The optimal protocol is therefore composed of intervals where $k=k_+$ and intervals where $k=k_-$ (``bang-bang'' protocol~\cite{chen_fast_2010,hegerfeldt_driving_2013,hegerfeldt_high-speed_2014,lu_fast_2014,ding_smooth_2020,prados_optimizing_2021,ruiz-pino_optimal_2022,patron_thermal_2022}), with the ``switchings'' between these values taking place at the times for which $\partial H/\partial k= 0$. This kind of protocols often appear as solution candidates for minimal-time problems
~\cite{hegerfeldt_driving_2013,hegerfeldt_high-speed_2014,prados_optimizing_2021,ruiz-pino_optimal_2022,patron_thermal_2022}, although a rigorous proof that it is the solution of the minimum time problem is only available when the Hamiltonian is linear in both $\uv$ and the control---which is not the present case, since $M_k$ depends on $k$.

In what follows, consistently with the notation introduced in the main text, the solution for the minimal time is denoted by $s_{\text{th}}$. The minimum number of switchings between $k_+$ and $k_-$ that we need, in order to match the boundary conditions, can be computed in this way: of the 6 conditions provided by the boundary conditions on $\uv$, 3 can be matched by the additive constants of the ordinary differential equations~\eqref{eq:dynamics-u}, while the remaining 3 need the introduction of an equal number of additional degrees of freedom (two switching times, $s'$ and $s''$, and the final time $s_{\text{th}}$). In other words, the simplest protocol fulfilling the requirements has the form
\begin{equation}
\label{eq:opt}
    k(t)=\begin{cases}
k_{+}&\quad 0\le s<s',\\
k_{-},&\quad s'\le s<s'',\\
k_{+}&\quad s'',\le s<s_\text{th},\\
    \end{cases}
    \quad \quad \text{ if } k_f>k_i\,.
\end{equation}
If $k_f<k_i$ we need to switch $k_+$ and $k_-$.
The switching times can be found by noticing that, for fixed values of $k=k_{\pm}$, the dynamics is linear in the translated variables
\begin{equation}
\uv_{\pm}  = \uv - \vect{r}_{\pm} \quad \text{ with } \quad \vect{r}_{\pm}=\begin{pmatrix}
    1/k_{\pm}\\ 0 \\ 1
\end{pmatrix}\,.
\end{equation}
The final state corresponding to the protocol~(\ref{eq:opt}) for 
$
\text{sgn}(k_f-k_i)=\pm 1
$
can be then explicitly computed as a function of $s'$, $s''$ and $s_{\text{th}}$:
\begin{subequations}
\begin{eqnarray}
    \uv (s') &=& \vect{r}_{\pm} + \exp\sbr{ s' M_{k_{\pm}}}\cbr{\uv_i-\vect{r}_{\pm}},\\
    \uv (s'') &=& \vect{r}_{\mp} + \exp\sbr{ (s''-s') M_{k_{\mp}}}\cbr{\uv(s')-\vect{r}_{\mp}},\\
    \uv (s_{\text{th}}) &=&  \vect{r}_{\pm} + \exp\sbr{ (s_{\text{th}}-s'') M_{k_{\pm}}}\cbr{\uv(s'')-\vect{r}_{\pm}}\,.    
\end{eqnarray}
\end{subequations}
By imposing
$
\uv(s_{\text{th}})=\uv_f,
$
one has 3 equations to be solved for $s'$, $s''$ and $s_{\text{th}}$: the solution can be obtained numerically.
The value of $s_{\text{th}}$ found in this way is the threshold for the protocol to exist. Its dependence on the boundary conditions is explored numerically, for fixed  $k_-$ and $k_+$, in Fig.~\ref{fig:mintime}.

Here, we have only considered three-step protocols of the form~(\ref{eq:opt}), but solutions with a larger number of switchings are in principle possible. From a physical point of view, one expects the solution with the minimum number of switchings to provide the minimum connection time. In fact, this has been proved in other non-linear control problems~\cite{prados_optimizing_2021,ruiz-pino_optimal_2022}, although we do not have a rigorous proof for our case.

\section{Minimal work}\label{sec:min-work}

\subsection{Derivation of the cost function for work minimization}
\label{sec:cost}

We want to compute an explicit expression for
\begin{equation}
\label{eq:costdef}
    {W(t_f)}=\int_0^{t_f}dt \, P(t)=\int_0^{t_f}dt \int_{-\infty}^{\infty}dx  \int_{-\infty}^{\infty}dv\, p(x,v,t) \, \partial_t U(x,t),
\end{equation}
where $U(x,t)$ is the time-dependent external potential, and we denote by $P(t)$ the (average) instantaneous power. Even if we are only interested in the particular case
\begin{equation}
U(x,t)=\frac{\kappa(t)}{2}x^2,
\end{equation}
it is useful to consider the problem in its full generality. See also~\cite{muratore-ginanneschi_extremals_2014} for an alternative derivation.

An integration by parts leads to
\begin{equation}
    P(t)=\partial_t \av{U} - \int_{-\infty}^{\infty}dx  \int_{-\infty}^{\infty}dv\, U(x,t) \,\partial_t p(x,v,t)\,,
\end{equation}
hence by recalling the Klein-Kramers equation
\begin{equation}
\label{eq:kram}
    \partial_t p(x,v,t)=-\partial_x \sbr{v  p(x,v,t)}+\frac{1}{m} \partial_v \cbr{\partial_x U(x,t)  p(x,v,t)+ \gamma v p(x,v,t)+\frac{\gamma k_B T}{m}\partial_v p(x,v,t)},
\end{equation}
we get
\begin{equation}
\label{eq:P1}
    P(t)=\partial_t \av{U} - \av{v \, \partial_x U(x,t) }\,.
\end{equation}
We now multiply both sides of~\eqref{eq:kram} by $mv^2/2$ and integrate in $x$ and $v$, obtaining
\begin{equation}
\partial_t \av{\frac{m v^2}{2}}=-\av{v \, \partial_x U(x,t) } - \gamma \av{v^2}+\frac{\gamma}{m} k_B T\,. 
\end{equation}
Once inserted into~\eqref{eq:P1} the above relation yields
\begin{equation}
\begin{aligned}
P(t)&=\partial_t \av{U}+ \partial_t \av{\frac{mv^2}{2}} +\gamma \av{v^2}- \frac{\gamma}{m} k_B T =\partial_t \av{E}+ \gamma \av{v^2}- \frac{\gamma}{m} k_B T\,,
\end{aligned}
\end{equation}
where $\av{E}$ is the average mechanical energy of the particle. From~\eqref{eq:costdef}, one thus gets 
\begin{equation}
\label{eq:worken}
    {W}(t_f)=\av{E(t_f)}-\av{E(0)}+\underbrace{\gamma \int_0^{t_f}dt \,\av{v^2} - \frac{\gamma}{m} k_B T t_f}_{-Q(t_f)}\,,
\end{equation}
where we have identified the exchanged heat $Q(t_f)$. The minimization of the work, once the boundary conditions are fixed, is therefore equivalent to the minimization of the cost function
\begin{equation}\label{eq:cost-C}
    \mathcal{C}\equiv -\frac{Q(t_f)}{k_B T} = \gamma \int_0^{t_f}dt \,\frac{\av{v^2}}{k_B T} - \frac{\gamma}{m} t_f\,.
\end{equation}
Reasoning in the same way, we can as well relate the average instantaneous power to the average free energy
\begin{equation}
\av{F}=\av{E}-TS
\end{equation}
where the entropy $S$ is defined as
\begin{equation}
    S=-k_B\int_{-\infty}^{\infty}dx  \int_{-\infty}^{\infty}dv\, p(x,v,t) \ln p(x,v,t)\,.
\end{equation}
By repeated use of~\eqref{eq:kram} and integration by parts, one finds
\begin{equation}
    \partial_t S=\frac{k_B^2\gamma T}{m^2}\int_{-\infty}^{\infty}dx  \int_{-\infty}^{\infty}dv\, p(x,v,t) \cbr{\partial_v \ln p(x,v,t)}^2-\frac{k_B \gamma}{m}\,.
\end{equation}
Equation~(\ref{eq:worken}) can thus be recast as
\begin{equation}
\begin{aligned}
    {W}(t_f)&=\av{F(t_f)}-\av{F(0)}+\gamma \int_0^{t_f} dt \int_{-\infty}^{\infty}dx  \int_{-\infty}^{\infty}dv\, \cbr{v+\frac{k_B T}{m}\partial_v \ln p(x,v,t)}^2 p(x,v,t)\\
    &=\av{F(t_f)}-\av{F(0)}+{W_{\text{irr}}}(t_f)\,.    
\end{aligned}
\end{equation}
Consistently with the second principle, the irreversible part ${W_{\text{irr}}}$ of the average work is always non-negative, and vanishes only at equilibrium. The minimal work is therefore attained along a quasistatic processes, where the instantaneous distribution is always of the Maxwell-Boltzmann type.

\subsection{Structure of the minimal work solution}

Our starting point is the cost function in Eq.~\eqref{eq:cost-C}, which in dimensionless variables reads
\begin{equation}
\mathcal{C}= \int_0^{s_f} ds\, \left(\z-1\right).
\end{equation}
The ``Lagrangian'' for this problem is $\mathcal{L}=\z-1$ and Pontryagin's Hamiltonian results
\begin{equation}
\begin{aligned}
    \mathcal{H}=-\z+1+\pv \cdot \cbr{M_k \uv + 2 \vect{e}_3},
\end{aligned}
\end{equation}
where $\pv$ is the vector of the moments conjugated to $\uv$, normalized in such a way that  $\pzero=-1$.

We start searching for the maximum of the Hamiltonian as a function of $k$, imposing $\partial\mathcal{H}/\partial k=0$, which gives
\begin{equation}
\label{eq:hw1}
    \w \py + 2 \y \pz=0\,.
\end{equation}
As we did before for the minimal time case, we consider the time derivative of the above identity, leading to
\begin{equation}
\label{eq:hw2}
\w \pw-\z \pz =\y\,,
\end{equation}
once the conditions~\eqref{eq:symdyn} are taken into account. In turn, the time derivative of~\eqref{eq:hw2} implies
\begin{equation}
\label{eq:hw3}
    2 \y \pw + \z \py -2 \pz=-k \,\w - \y +2 \z \,.
\end{equation}
At variance with the case considered before, now the algebraic linear system \eqref{eq:hw1}--\eqref{eq:hw3} for $\pv$ is inhomogeneous, and therefore $\pv^T\ne(0,0,0)$. By taking the derivative with respect to $s$ once again, making repeated use of~\eqref{eq:symdyn} and bringing to bear Eqs.~(\ref{eq:hw1})--(\ref{eq:hw2}), one finally concludes that
\begin{equation}
\label{eq:ksol}
    \frac{d k}{d s}=3k+\frac{6}{\w}+3\frac{\y}{\w}-9\frac{\z}{\w}+10\frac{\y^2}{\w^2}-8\frac{\y \z}{\w^2}+8\frac{\y^3}{\w^3}\,.
\end{equation}

Equation~\eqref{eq:ksol}, coupled with the dynamics of $\uv$, provides a fourth-order differential system. Let us call $k_0(s)$ its solution, specified up to four arbitrary constants (degrees of freedom). This function $k_0(s)$ alone, with our boundary conditions, cannot be the solution of our problem---it does not have enough degrees of freedom to match the six boundary conditions: the initial and final values of the three second moments. Note that this difficulty cannot be solved by introducing Dirac-delta jumps in the stiffness at the initial and final times, because of the stiffness being bounded in the interval $0\leq k_-\leq k\leq k_+$.

Reasoning as we did for the minimal-time case, and taking into account the fact that now the final time $s_f$ is fixed, we realize that the simplest possible solution satisfying all the boundary conditions is obtained by alternating two intervals with fixed $k=k_{\pm}$ (from $s=0$ to $s=s_1$ and from $s=s_2$ to $s=s_f$) with an interval (from $s_1$ to $s_2$) with $k=k_0(s)$. This three-stage solution, which has been analyzed in the main text, has six degrees of freedom: the four arbitrary constants in $k_0(s)$ plus the two switching times $s_1$ and $s_2$. 

\new{The mentioned solution (including the values of $s_1$ and $s_2$) can be found numerically.
We first notice that the values of $u_1$, $u_2$ and $u_3$ at $s_1$ and $s_2$ are completely determined, once $s_1$ and $s_2$ are known: one just needs to evolve the boundary conditions at $s=0$ ($s=s_f$) forward (backward) for a time $s_1$ ($s_f-s_2$), with the fixed stiffness $k_{\pm}$, depending on the compression/decompression nature of the problem. We are thus left with a 4-th order ordinary differential system in the $[s_1,s_2]$ interval, whose boundary conditions and integration time depend on the unknown parameters $s_1$ and $s_2$. The idea is to map this problem into a 6-th order differential system, where $s_1$ and $s_2$ are variables, and solve it numerically with a standard shooting method.}

\new{To this aim, the first step is to rewrite the equations for the dynamics and Eq.~\eqref{eq:ksol} in terms of a rescaled time coordinate
\begin{equation}
    z=\frac{s-s_1}{s_2-s_1}\,.
\end{equation}
We get:
    \begin{subequations}
\label{eq:rescz}
\begin{eqnarray}
         \frac{du_1}{dz}&=&(s_2-s_1)\cbr{2 u_2}\\
         \frac{du_2}{dz}&=&(s_2-s_1)\cbr{-k u_1 -u_2+u_3}\\
        \frac{du_3}{dz}&=&(s_2-s_1)\cbr{-2k u_2 -2 u_3+2}\\
        \frac{d k}{d z}&=&(s_2-s_1)\cbr{3k+\frac{6}{\w}+3\frac{\y}{\w}-9\frac{\z}{\w}+10\frac{\y^2}{\w^2}-8\frac{\y \z}{\w^2}+8\frac{\y^3}{\w^3}}\,.   
\end{eqnarray}
    \end{subequations}
In this way the boundary conditions at $z=0$ and $z=1$ (i.e., $s=s_1$ and $s=s_2$) still depend upon $s_1$ and $s_2$, but the integration time does not.
}
\new{Now we change the parameters $s_1$ and $s_2$ into variables of the system, by introducing two additional equations, namely
    \begin{subequations}
\label{eq:addeq}
\begin{eqnarray}
        \frac{ds_1}{ds}&=&0\\
        \frac{ds_2}{ds}&=&0\,.    
\end{eqnarray}
    \end{subequations}
These, once coupled with~\eqref{eq:rescz}, form a 6-th order differential system, which can be solved numerically by standard ``shooting'' methods. It is interesting to notice that even if the boundary conditions on $u_1$, $u_2$ and $u_3$ depend on $s_1$ and $s_2$, a shooting strategy is still able to find a solution: the method involves indeed a guessing of the initial values for the unknown variables $k$, $s_1$ and $s_2$ at $z=0$, which also fixes the boundary conditions for $\vect{u}$.
}

It is important to stress the differences of both our approach and our results with those in Ref.~\cite{gomez-marin_optimal_2008}. Therein, the minimization of the work for a harmonically confined underdamped particle was also considered, but there are two key differences. First, the stiffness had no bounds, i.e., it was admissible to have any real value for $k$, $-\infty<k<+\infty$. Second, the boundary conditions were different, the target state was not enforced to be the equilibrium state corresponding to the target value of the stiffness. 

Due to the absence of bounds in the stiffness, the work minimization in Ref.~\cite{gomez-marin_optimal_2008} was carried out by a variational calculus approach. As stated in Sec.~\ref{sec:pontryagin}, the variational approach is equivalent to Pontryagin's when the maximum of the Hamiltonian stems from the condition $\partial\mathcal{H}/\partial k=0$. Therefore, the variational calculus framework led to a fourth-order differential system, equivalent to Eq.~\eqref{eq:ksol} plus the evolution equations for $\uv$, i.e., to the solution $k_0$. The mismatch between the number of degrees of freedom in $k_0$ (four) and the number of boundary conditions (also six, while different from ours) was solved by introducing Dirac-delta peaks at the initial and final times---due to the unbounded stiffness. 

In the limit $\{k_-\to-\infty,k_+\to+\infty\}$, our three-stage solution for bounded stiffness tends to a solution with Dirac-delta peaks at the initial and final times, and $k_0(s)$ for $0<s<s_f$---because $s_1\to 0^+$ and $s_2\to s_f^-$. Yet, it is worth stressing that, due to the different boundary conditions, this limit of our three-stage solution is not the protocol found in Ref.~\cite{gomez-marin_optimal_2008}: the final state in the swift state-to-state transformation (SST) developed in Ref.~\cite{gomez-marin_optimal_2008} would evolve in time for times longer than $s_f$, whereas our system remains stationary for $s>s_f$.

\section{Optimal work for $s_f$ close to $s_{\text{th}}$}
\label{sec:shorttimes}

In the main text, we argue that, if the total time $s_f$ is fairly larger than the threshold $s_{\text{th}}$, a minimal-work solution can be found in the form of a three-stage protocol. The first and the last steps are performed at constant $k=k_{\pm}$ (the sign depends on the case), while the intermediate stage is  given by the solution $k_0(s)$ of the Euler-Lagrange equations. In this section, we show some explicit numerical examples to gain insight into the case of $s_f$ close to $s_{\text{th}}$.

\begin{figure}[h!]
\centering
\includegraphics[width=.95\textwidth]{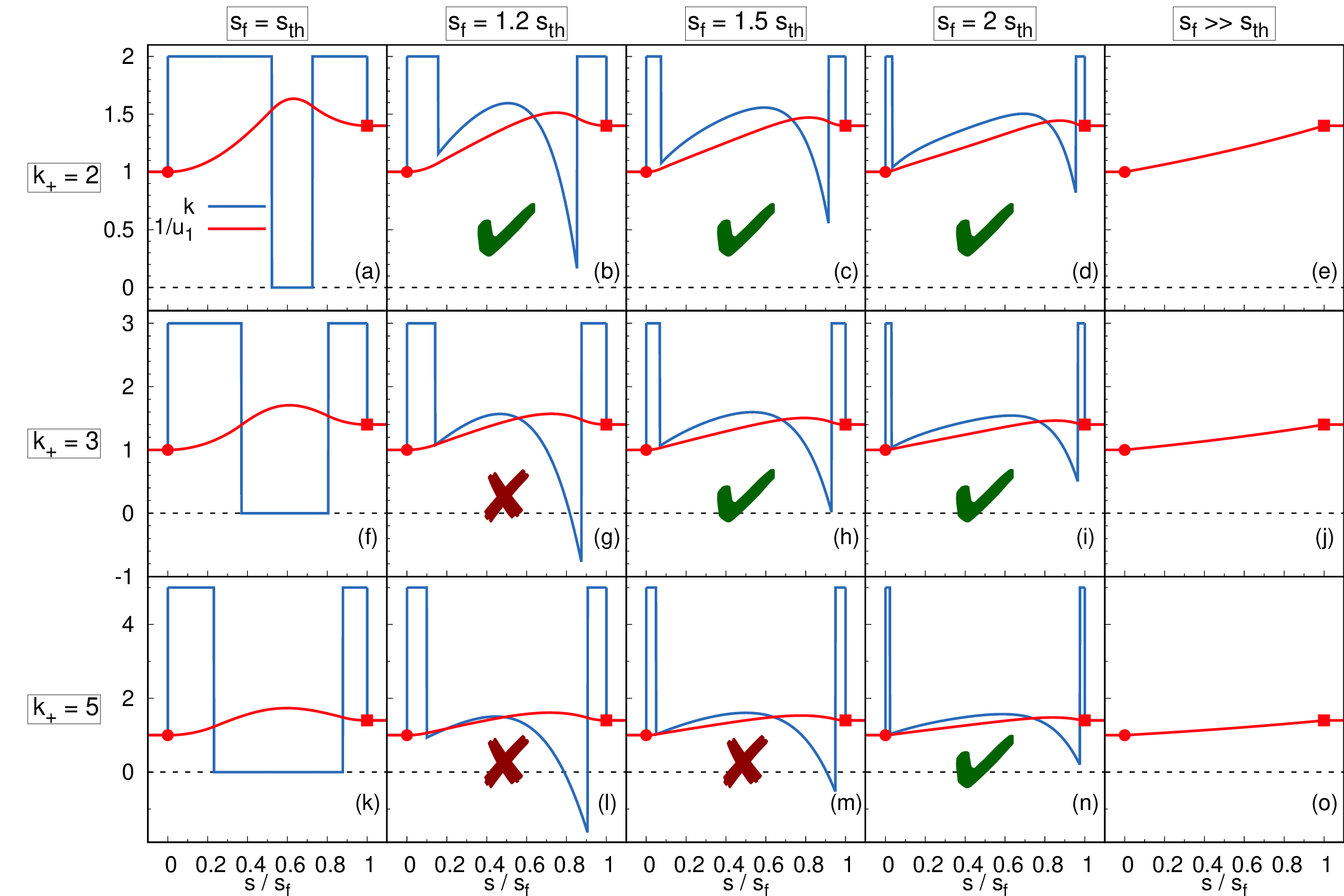}
\caption{Minimal work protocols. In each panel the optimal protocol $k$ (blue) and the inverse of the position variance $1/u_1$ (red) are shown. The leftmost column [panels (a), (f) and (k)] shows the three-step minimal time protocol, which takes place by definition in a rescaled time $s_{\text{th}}$. We expect this process to be the only possible one for $s_f=s_{\text{th}}$, given the bounds $k_-\le k \le k_+$, and therefore to be also the minimal-work process for that time horizon. The second, third and fourth column show the (tentative) minimal-work three-stage protocols for $s_f=1.2 s_{\text{th}}$,  $s_f=1.5 s_{\text{th}}$ and  $s_f=2 s_{\text{th}}$. Different rows correspond to different values of $k_+$, for fixed $k_-=0$. We notice that the prescribed recipe to build the minimal work protocol fails in the cases of panels (g), (l) and (m), as the protocol found by solving the differential system for $s_1<s<s_2$ exceeds the bounds. \new{The rightmost column [panels (e), (j) and (o)] shows the limit of quasi-static protocol, in which $k$ and $1/u_1$ coincide.} Here $k_i=1$, $k_f=1.4$. 
}
\label{fig:param}
\end{figure}

In Fig.~\ref{fig:param} we show the explicit three-stage solution for different values of the total time $s_f$ and of the upper bound $s_+$, with $k_-=0$. As expected, when $s_f$ is long enough, the solution built in this way is admissible, meaning that $k_0(s)$ never exceeds the interval $[k_-,k_+]$. However, as $s_f$ approaches $s_{\text{th}}$, the optimal solution tends to span a wider range of values for the stiffness, eventually exceeding the bounds on $k$---see panels (g), (l) and (m) of Fig.~\ref{fig:param}.

Since $s_f> s_{\text{th}}$, we know that completing a transition between the prescribed initial and final states must be possible. We also know that, among the many possible solutions, the one requiring minimal average work must satisfy Pontryagin's maximum principle. We may therefore search for four-stage protocols of the form
\begin{equation}
\label{eq:opw4}
    k(s)=\begin{cases}
k_{\pm},&\quad 0\le s<s_1,\\
k_{0}(s),&\quad s_1\le s<s^{\star},\\
k_{\mp},&\quad s^{\star}\le s<s_2,\\
k_{\pm},&\quad s_2\le s<s_{f},\\
    \end{cases} \quad \quad \quad \text{or}\quad \quad \quad
    k(s)=\begin{cases}
k_{\pm},&\quad 0\le s<s_1,\\
k_{\mp},&\quad s_1\le s<s^{\star},\\
k_{0}(s),&\quad s^{\star}\le s<t_2,\\
k_{\pm},&\quad s_2\le s<t_{f}\,,\\
    \end{cases}
\end{equation}
with the initial sign depending on the boundary conditions. Both forms are in principle allowed by Pontryagin's condition: one can discriminate between the two on the basis of a visual inspection of the unbounded solution, and invoking a continuity argument. The time $s^{\star}$ needs to be found by numerically minimizing the cost function.

\begin{figure}[h!]
\centering
\includegraphics[width=.9\textwidth]{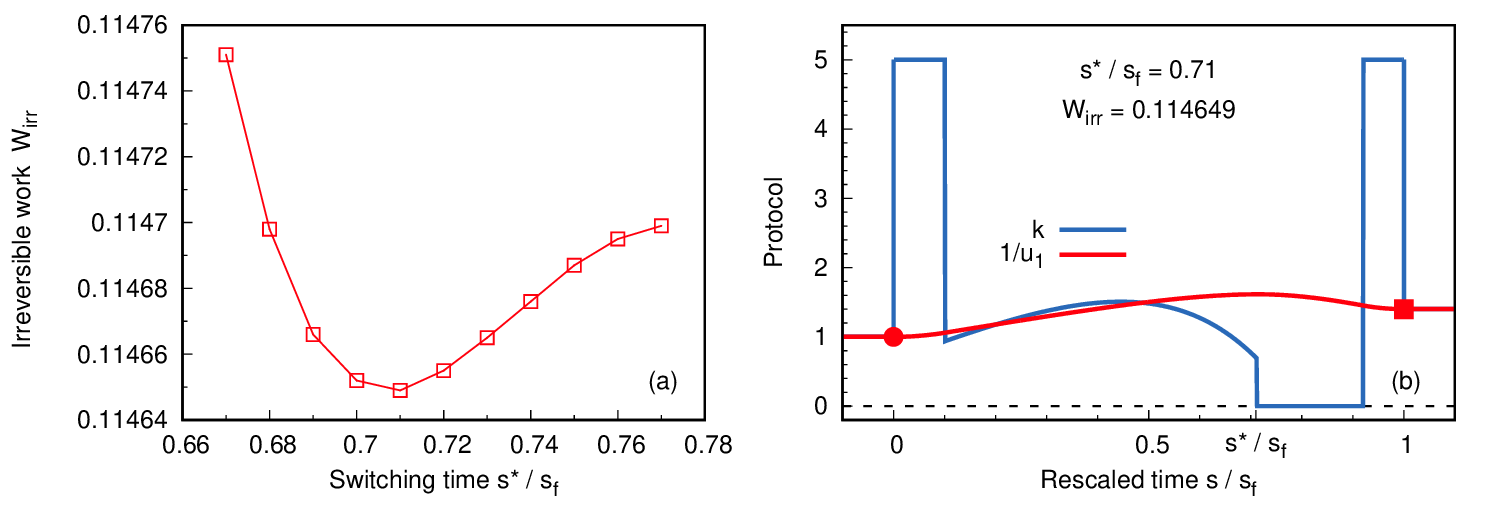}
\caption{Selection of the optimal four-stage protocol. Panel (a) shows the value of the average irreversible work, in dimensionless units, as a function of the switching time $s^{\star}$ appearing in~\eqref{eq:opw4}. By inspection of this curve, one can select empirically the value corresponding to the minimal work (here $s^{\star}\simeq 0.71 s_f$). In panel (b), the corresponding protocol is shown, with the same color code as in Fig.~\ref{fig:param}. The parameters are the same of panel (l) of Fig.~\ref{fig:param}. 
}
\label{fig:analysis4}
\end{figure}

Here we consider the explicit example of panel (l) of Fig.~\ref{fig:param}. Since $k_0$, in the three-stage solution, gets lower than $k_-$ just before the switching time $s_2$, we expect the first of the protocols~\eqref{eq:opw4} to be the correct one, by continuity. We thus compute the work obtained by such protocols for different values of $t^{\star}$: the results of this analysis are shown in Fig.~\ref{fig:analysis4}(a). We observe that the irreversible work attains a minimum, which corresponds therefore to the optimal protocol (at least, among the four-stage protocols that we are considering). Its explicit form is shown in  Fig.~\ref{fig:analysis4}(b).

\begin{figure}[h!]
\centering
\includegraphics[width=.95\textwidth]{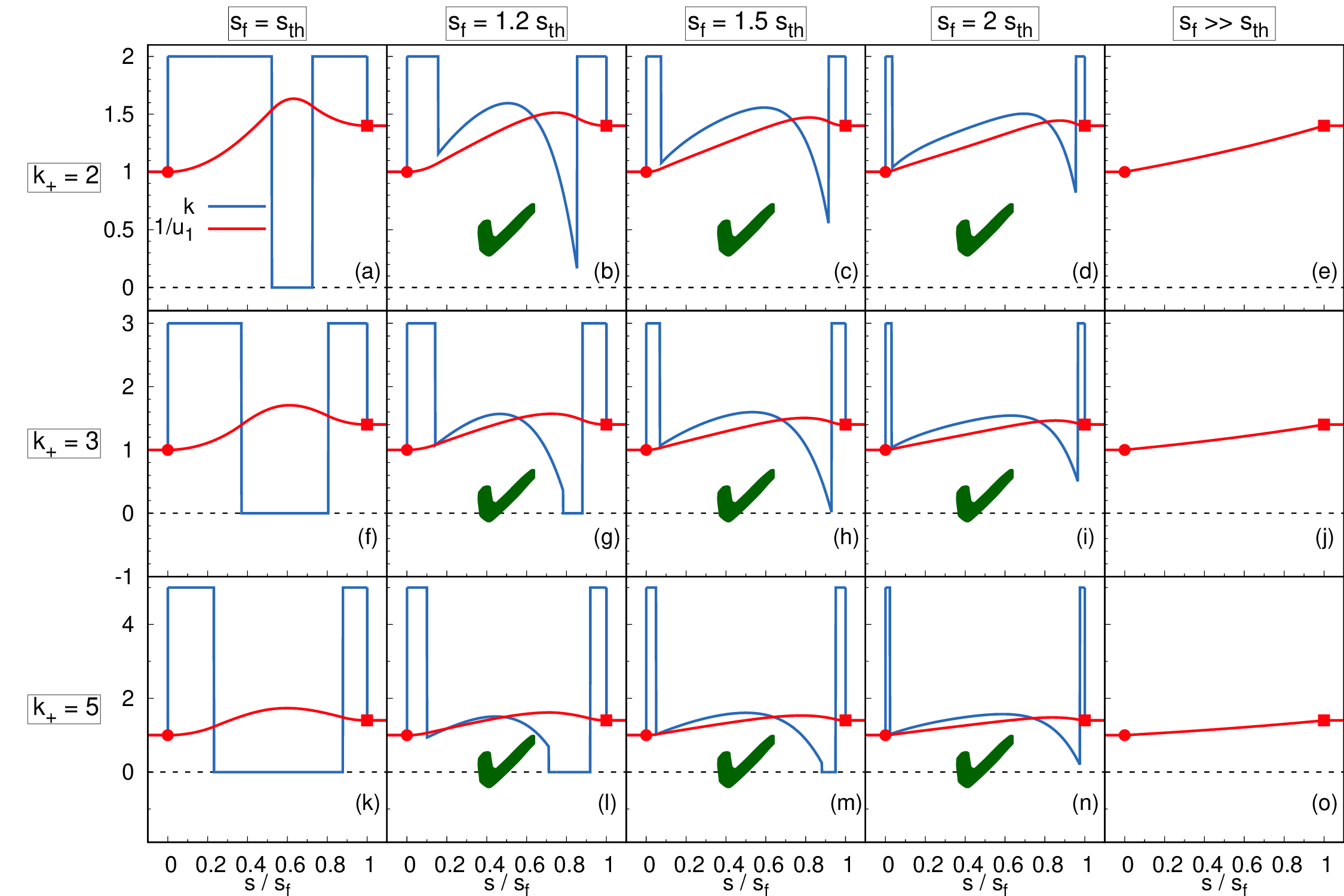}
\caption{Same as in Fig.~\ref{fig:param}, where panels (g), (l) and (m) have been replaced by four-stage protocols obtained as shown in Fig.~\ref{fig:analysis4}. 
}
\label{fig:param_amended}
\end{figure}

A similar reasoning can be applied to every non-admissible plot in Fig~\ref{fig:param}. We obtain in this way the ``amended''  Fig~\ref{fig:param_amended}. For values of $s_f$ even closer to $s_{\text{th}}$, more complex protocols may arise. The study of these quite specific situations is out of the scope of the present paper.

\section{Calibration of experimental trajectories time traces}
\label{sec:calibration}
The particle position $x(t)$ is experimentally measured using a photodetector. The recorded voltage writes 
\begin{equation}
    V_x(t) = c_x x(t) + N(t)\, , 
\end{equation}
where $c_x$ is the position calibration factor, and $N$ is the experimental noise assumed to be uncorrelated with the particle displacement. Consequently, the variance of the measured signal writes 
\begin{equation}
    \sigma_V^2=c_x^2\sigma_x^2(t) + \sigma_N^2\, .
\end{equation}

One can show that in the case of the  STEP protocol, the particle position variance is given by~\cite{raynal_shortcuts_2023}
\begin{equation}
    \sigma^2_x(t)= \sigma_i^2 \frac{\chi-1}{2\chi}[1+\cos(2\Omega_f t)]e^{-\gamma t}+\frac{\sigma_i^2}{\chi} \, .
    \label{eq:step_var}
\end{equation}
Experimentally, we enforce the compression factor $\chi=0.5$, the trap frequency $\Omega_f$, i.e., $\kappa_f$, and the initial and final position variances through the equipartition theorem
\begin{align}
    \sigma_i^2 = \frac{k_B T}{\kappa_i}, \qquad
    \sigma_f^2 = \frac{k_B T}{\kappa_f} .
\end{align}
Thus, from a fit of the experimental STEP protocols, we can retrieve the values of $c_x$ and $\sigma_N$ to determine the actual particle position variance.



\section{Computation of the mean experimental work}
\label{sec:computation}
The cumulative work up to time $t$ is given by
\begin{equation}
\label{eq:work}
  W(t) = \int_0^{t}dt'\,\frac{\dot{\kappa}(t')}{2}  \av{x^2(t')}\,.
\end{equation}
For protocols without stiffness discontinuity (SST, or Euler-Lagrange part of the bounded optimal protocols), we can compute numerically the integral~\eqref{eq:work}, 
where we use the theoretical value of $\kappa$ and the experimentally measured value of $\sigma_x$. For protocols with stiffness discontinuity at time $t_d$, the associated instantaneous work is
\begin{equation}
      W_d(t_d) = \frac{1}{2}\left[\kappa(t_d^+)-\kappa(t_d^-)\right]\langle x^2(t_d^-) \rangle\, . 
\end{equation}
To minimize the impact of the exact choice of  $t_d^-$ on the computation, we oversample the experimental data to get a smaller time step ($\delta t_\text{W, oversampled}=1.6$~ns, and $\delta t_\text{acquisition}=200$~ns). We note that this oversampling neither affects, beyond measurement uncertainties, the computation of the work on the Euler-Lagrange segment of the optimal protocols nor on the SST protocol.

In the case of the STEP protocols, since they have been used to calibrate our measurement scheme by construction, the associated work is given by 
\begin{equation}
     W_\text{STEP}= \frac{1}{2}(\kappa_f-\kappa_i)\sigma_i^2
      = \frac{k_B T}{2}(\chi-1) 
      = -\frac{k_B T}{4}   , 
\end{equation}
for our choice $\chi=0.5$.

Note that we assumed that the discontinuities are perfectly stiff, since the typical time required to change the stiffness is $\tau_\text{raise}\approx 200$~ns, much shorter than all other timescales of the problem.



\bibliography{biblio, Mi-biblioteca-24-abr-2025}